# Femtosecond electron diffraction reveals local disorder and local anharmonicity in thermoelectric SnSe

Jingjun Li[1#], Yingpeng Qi[1*], Qing Yang[2#], Luye Yue[1], Changyuan Yao[1], Zijing Chen[3,4], Sheng Meng[2,5,6], Dao Xiang[3,4,7,8], Jianming Cao[9,1]

[1]Center for Ultrafast Science and Technology, School of Physics and Astronomy, Shanghai Jiao Tong University, Shanghai 200240, China.

[2]Beijing National Laboratory for Condensed Matter Physics and Institute of Physics, Chinese Academy of Sciences, Beijing 100190, China.

[3]Key Laboratory for Laser Plasmas (Ministry of Education), School of Physics and Astronomy, Shanghai Jiao Tong University, Shanghai 200240, China.

[4]Collaborative Innovation Center of IFSA (CICIFSA), Shanghai Jiao Tong University, Shanghai 200240, China.

[5]School of Physical Sciences, University of Chinese Academy of Sciences, Beijing 100049, China.

[6]Songshan Lake Materials Laboratory, Dongguan, Guangdong 523808, China.

[7]Tsung-Dao Lee Institute, Shanghai Jiao Tong University, Shanghai 200240, China.

[8]Zhangjiang Institute for Advanced Study, Shanghai Jiao Tong University, Shanghai 200240, China.

[9]Physics Department and National High Magnetic Field Laboratory, Florida State University, Florida 32310, United States of America.

#Contributed equally

*Correspondence to qiyp@sjtu.edu.cn

## Abstract

The microscopic arrangement of atoms and molecules is the determining factor in how materials behave and perform. Beyond the long-range periodicity, the local disorder with local structures deviating from the average lattice structure plays a vital role in determining the physical properties of the phonon, electron and spin subsystems in crystalline functional materials. Experimentally characterizing the 3D atomic configuration of such local disorder and correlating it with the advanced functions remain a big challenge. Time-domain evolution of the local disorder, either static or dynamical, is lost due to the characterization at equilibrium state with conventional probing techniques. With the combination of femtosecond electron diffraction, structure factor calculation and TDDFT-MD simulation, we exclusively identify the static local disorder and the local anharmonicity of it in thermoelectric SnSe. The ultrafast structural dynamics in time domain reveal a dominant static off-symmetry displacement of Sn (~0.4 Å) in the ground state and the anharmonicity of this local disorder induces an ultrafast atomic displacement within 100 fs after photoexcitation. The microscopic picture of the local anharmonicity indicates a direct and first signature of the THz Einstein oscillators in real space. Therefore, a glass-like thermal transport channel with the local disorder, the Einstein oscillators and the local anharmonicity, updates the fundamental insight into the long-debated ultralow thermal conductivity in SnSe. The local disorder over one to a few unit cells is pervasive and indispensable in thermoelectric materials, multiferroic materials and correlated electronic materials. Our method of revealing the 3D local disorder and the local correlated interactions by ultrafast structural dynamics will inspire broad interest in construction of the structure-property relationship in material science.

## 1. Introduction

That structure determines the property is a traditional paradigm in materials science. Understanding the intrinsic correlation between structure and functional properties lies at the heart of materials science and engineering. Crystal structure based on long-range periodicity provides excellent convenience in description and coherent control of crystalline properties [1-4]. However, emerging functional crystalline materials usually harbor local disorder deviating from the global long-range average structure [5]. Critically, such local disorder has a profound impact on their functional properties of the phonon, electron and spin subsystems, such as the correlated electronic state [6-8], the thermal conductivity [9, 10], and the structural transition [11-13]. From a conventional crystallographic viewpoint, there is no distinction between the crystalline materials with and without the local disorder, since their configurational averages are characterized by the same crystallographic space group. Though much effort has been devoted to determine the local disorder in such crystalline disordered materials [10, 13-17], a quantitative 3D characterization of the local atomic configuration and the local potential energy environment is still a big challenge. An open question is whether the local atomic configuration is dominated by local correlated interactions or in a random manner [18]. In this work, we propose using femtosecond electron diffraction to detect the non-equilibrium structural dynamics in crystalline disordered materials and disentangle the intrinsic local disorder from the average lattice structure with femtosecond-picometer temporospatial resolution. Specifically, with the combination of femtosecond electron diffraction, structure factor calculation and the TDDFT-MD simulation, we exclusively identify the static local disorder and the anharmonicity of this local disorder in thermoelectric SnSe. As shown in Fig. 1a, with ultrafast laser-pump electron-probe, we disentangle the local disorder and local anharmonicity from the global symmetry and global anharmonicity. The local disorder with 8 degenerated atomic configurations evidences correlated local interactions in crystalline disorder materials. A glass-like thermal transport channel with the local disorder, the Einstein oscillators and the local anharmonicity, is determined and updates the fundamental insight into the long-debated ultralow thermal conductivity in SnSe.

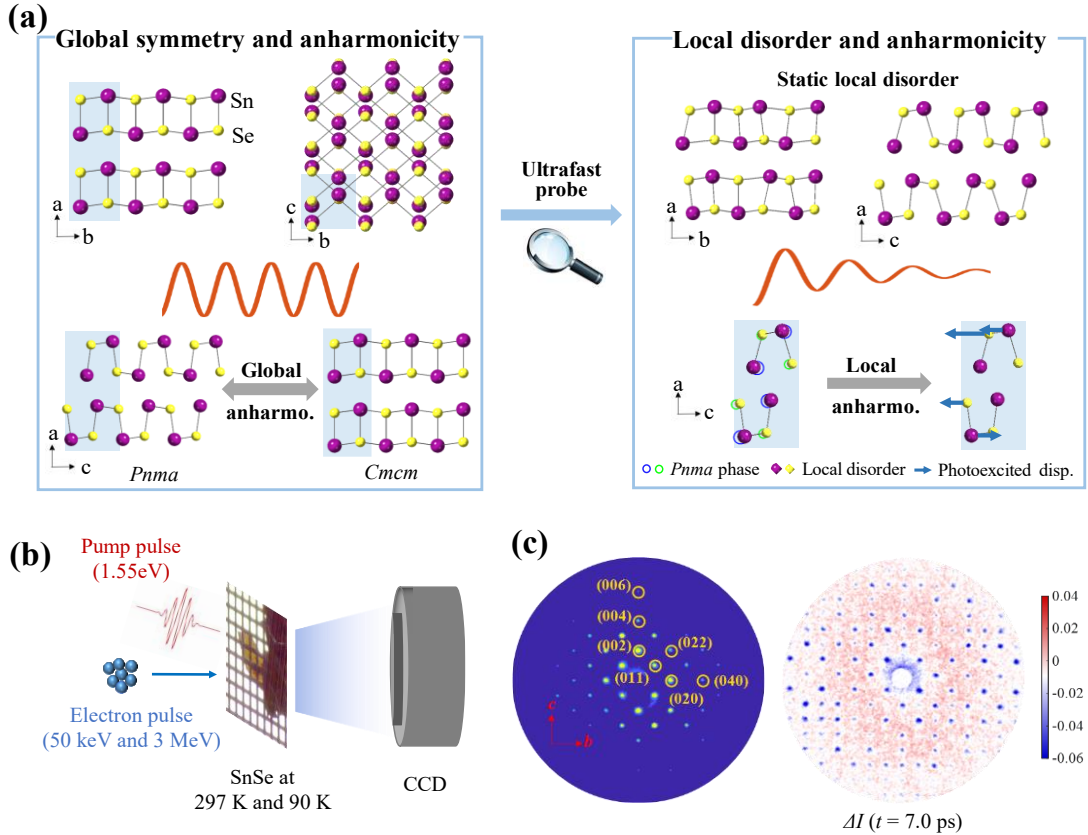

**Fig. 1. Probing the static local disorder and the local anharmonicity in SnSe by femtosecond electron diffraction. (a)** Left: Crystal structure of SnSe in *Pnma* phase with the side view (*a-b* and *a-c* plane) and the top view (*b-c* plane); The size of the unit cell is labeled by light blue rectangles; The oscillation curve in the middle panel indicates a long-range coherence from the global symmetry; The bottom panel displays the structural transition from the *Pnma* to *Cmcm* phase as rising temperature dominated by a global anharmonicity. Right: the top panel is the schematic of the revealed static local disorder along the *b* axis and the *c* axis with the dominant off-symmetry displacements of Sn atoms among unit cells; such local disorder reduces the long-range coherence to a damped oscillation curve in the middle panel; the bottom panel displays the photoexcited displacements of Sn and Se atoms determined by the local anharmonicity; the hollow circles and the solid circles represent the rigid crystal structure and the local disordered structure respectively. **(b)** Schematic of the ultrafast pump-probe experiments with 50 keV and 3 MeV femtosecond electron diffraction setup. The SnSe nanofilm on TEM grid is pumped by 35 fs, 800 nm laser pulse at 297 K and 90 K, and the diffracted patterns at different time delays are recorded by a CCD camera. **(c)** Left: Static diffraction pattern acquired during experiments. Several Bragg reflections are labeled by Miller index. Right: Differential pattern at 7.0 ps after photoexcitation.

Thermoelectric materials, based on the direct conversion between thermal and electrical energy, have been considered as promising alternatives to meet the challenges of the global energy dilemma [19, 20]. Single crystalline SnSe has drawn strong attention recently in the thermoelectric community due to the surprising ultralow thermal conductivity ($\kappa_{lat}$ < 1 Wm$^{-1}$ K$^{-1}$) and extraordinarily high thermoelectric figure of merit ($ZT$ > 2) [21-23] in such a simple binary compound. The crystal structure of layered SnSe, as displayed in the left figure of Fig. 1a, holds a puckered distortion along the $c$ axis and a thermal-induced structural transition from the *Pnma* to *Cmcm* phase at ~807 K [22, 24]. The physical mechanism of the ultralow thermal conductivity at both the ambient temperature and the high temperature remains controversy. The phonon mean free path of SnSe at room temperature is ~0.84 nm [25], comparable to the lattice parameters of the unit cell in *Pnma* phase. With inelastic neutron scattering measurements, the ultralow thermal conductivity is qualitatively attributed to lattice anharmonicity acrossing the structural transition (*Pnma* to *Cmcm* phase) [24, 26]. However, the recent experiments show that the anharmonic behavior starts near room temperature [27, 28], substantially lower than the structural transition temperature. The relatively flat thermal conductivity without abrupt decrease at high temperature also challenge the lattice anharmonicity enhanced by the structural transition. Besides the lattice anharmonicity, the vacancies and off-stoichiometry in the ground state of SnSe are identified [29-31]. Most SnSe sample in research is intrinsically p doping, deriving from the vacancy defects [22, 29]. These point defects and associated local structures may invalid the phonon picture and complicate the thermal transport channel with the Debye model (particlelike phonons in crystal), the Einstein model (localized oscillators in glass) and the anharmonicity in both models [17, 32-34]. Yet, conclusive and quantitative characterization of the 3D local structure and the thermal transport channel are not clear in SnSe.

Probing and characterizing the local structure in crystalline materials is a difficult experimental challenge. The conventional methods [10, 14-17, 35], such as the neutron total scattering, characterize the local structure at equilibrium state within the radial direction. Therefore, the time-domain evolution of the local disorder, either static or dynamical [13] with the underlying feature of the local potential energy landscape, is lost. To resolve this, we propose using femtosecond electron diffraction to disentangle in time domain the vibrational and displacive motion of atoms in local structure [11, 36, 37]. A schematic illustration of the femtosecond electron diffraction is shown in Fig. 1b. By varying the time delay between the pump laser and the electron probe, the transient structural dynamics is detected with the temporal resolution up to ~50 fs [38]. The acquired prototypical diffraction pattern and the differential pattern in experiments are displayed in Fig. 1c. Generally, distinct local structure from the average lattice structure induces anomalous intensity change of the Bragg reflections and the diffuse scattering [39]. The Bragg reflections have an intensity several order of magnitude larger than that of the diffuse scattering, so it potentially has higher signal to noise ratio and higher priority in quantifying the local structure [11, 36, 40]. The local structure has very limited correlation length over one to a few unit cells, the reflection intensity is thus determined by such a supercell [41, 42]. The intensity of a Bragg reflection, $I \propto |F|^2$, at the time delay point $\Delta t$ can be calculated by the structure factor:

$$F(hkl, \Delta t) = \sum_j T_j f_j \cdot exp[-i2\pi(h(x_j + \Delta x_{j,\text{diso}} + \Delta x_{j,\text{disp}}) + k(y_j + \Delta y_{j,\text{diso}} + \Delta y_{j,\text{disp}}) + l(z_j + +\Delta z_{j,\text{diso}} + \Delta z_{j,\text{disp}}))] \quad (1)$$

Where the summation runs over all atoms in the supercell of the local structure, $f_j$ is the atomic form factor for the $j$-th atom, $r_j = x_j \hat{a} + y_j \hat{b} + z_j \hat{c}$ is the vector position of the atom in the supercell and (*hkl*) is the Miller indices. $\Delta x_{j,\text{diso}}$, $\Delta y_{j,\text{diso}}$ and $\Delta z_{j,\text{diso}}$ are the displacements from the static disorder at the ground state. $\Delta x_{j,\text{disp}}$, $\Delta y_{j,\text{disp}}$ and $\Delta z_{j,\text{disp}}$ are the time-dependent displacements induced by ultrafast photoexcitation. $T_j$ is the thermal vibration, i.e., the Debye-Waller effect, induced by ultrafast photoexcitation:

$$T_j = exp(-M), \quad M = 8\pi^2 \langle \mu^2 \rangle (\sin\theta/\lambda)^2 \tag{2}$$

$M$ denotes the Debye-Waller factor, $\theta$ is half the scattering angle and $\langle \mu^2 \rangle$ represents the mean square displacement of atoms in the supercell. Therefore, the ultrafast structural dynamics in time domain enable the identification of the thermal vibration, the static displacement of the local disorder and the photoexcited displacement. The photoexcited displacement denotes the anharmonicity of the static local disorder. If the local structure is dynamical in time and space, the average over the repeated pump-probe cycles with kHz frequency gives rise to an overall Debye-Waller effect, distinct from that of the static local structure. For SnSe, the constrains from the transient structural dynamics of the total scattering along the crystallographic $b$ axis, $c$ axis, the diagonal direction and the ultrafast dynamics within 100 fs, exclusively determine the static local disorder with multiple atomic configurations and the anharmonicity of the local disorder with picometer precision.

## 2. Identification of the local disorder in *Pnma* phase of SnSe

SnSe nanofilms with the thickness of ~50 nm are exfoliated from the bulk crystal and characterized carefully as shown in Supplementary S1. With 800 nm femtosecond laser excitation, the threshold fluence, the photoexcited carrier density and the photoinduced temperature rising of SnSe are summarized in Supplementary S2. To investigate the ultrafast structure dynamics, we firstly focus on the time-resolved intensity changes of Bragg reflections because of the high signal to noise ratio. The crystallographic $b$ axis is the high symmetry axis, so the intensity changes of Bragg reflections (0$k$0) ($k$ = 2, 4, 6) along this axis is analyzed as the first step. In Fig. 2a, the intensity of these reflections decay simultaneously with almost the same amplitude after photoexcitation. The temporal evolution is well fitted by a single exponential function with the time constant of ~3.9 ps, indicating a single physical process along the *b* axis. Photoexcitation enhanced thermal vibration, i.e., the Debye-Waller effect, is usually expected to dominant such behaviors of reflections along a high symmetry axis [36, 43, 44]. While, the calculated intensity changes with increasing the thermal vibration amplitude display a significantly larger intensity change for the higher order reflections than that of the lower order reflections, as shown in the inset of Fig. 1a and Supplementary S3.1. The Debye-Waller effect therefore cannot reproduce the experimental phenomenon. We exclude possible impacts from inhomogeneous laser pump and multiple scattering effect [45-47] (see Supplementary S3.2 and S3.3). Other possible contributions from arbitrary atomic displacements, photoinduced formation of polarons [41, 42], stacking faults [48, 49], their combination with Debye-Waller effect and photoinduced changes of the atomic form factor [50-52] are also excluded by comprehensive studies (see Supplementary S3.4 to S3.7). Therefore, beyond the rigid crystal structure of *Pnma* phase used in above analyses, the possible static local off-symmetry displacements induced by vacancies [29-31] need to be introduced in the ground state of SnSe.

To further analyze the structural dynamics along the *b* axis, the pump fluence dependent intensity change of the (020) reflection is shown in Fig. 2b. At varying time delays, the inset of Fig. 2b displays a linear increase of the intensity change amplitude as a function of the pump fluence, following a conventional character of Debye-Waller effect. A possible model is thereby the photoexcitation enhanced Debye-Waller effect in the ground state of SnSe with local off-symmetry displacements. Closing to the threshold fluence of ~9.5 mJ/cm$^2$, the intensity change of the (020) reflection is better

8 / 26

fitted by a bi-exponential function with the time constants of $\tau_1 = 3.9 \pm 0.7$ ps and $\tau_2 = 28.4 \pm 6.8$ ps, as shown in Fig. 2c. Such a two-step intensity change indicates potentially a fast electron-phonon coupling and a subsequent slow phonon-phonon coupling [53, 54]. The similar two-step intensity change also displays on the (040) reflection as shown in Fig. 2c. The two amplitudes of the bi-exponential fit ($a_1$ and $a_2$) for (020) reflection are respectively equal to that of the (040) reflection within the error range, as depicted in the inset of Fig. 2c. Therefore, a consistent structural dynamic, i.e., the Debye-Waller effect in the ground state of SnSe with static local off-symmetry displacements, dominates the intensity change over the whole-time scale. The intensity change of more other reflections in Supplementary S4 further confirms this structural dynamic.

To quantify the static local off-symmetry displacements and the Debye-Waller effect, we searched the best fit between the time-resolved experimental and calculated intensity changes over the Bragg reflections along the *b* axis (see details in Supplementary S3.8). The random substitution of Sn with vacancies induces potentially random local distortions within unit cells. The diffuse scattering in Supplementary S1 and S9 displays no signature of supercells, so the correlation length of the local disorder is within a single unit cell. The unit cell of SnSe in *Pnma* phase contains eight atoms. Eight static displacements for each atom ($\Delta y_{1\_disp}$ to $\Delta y_{8\_disp}$) and two thermal vibrations ($\mu_{Sn}$ and $\mu_{Se}$) are used in searching the best fit to experimental results at 60 ps. Then a monotonous increase of $\mu_{Sn}$ and $\mu_{Se}$, mimicking the experimental Debye-Waller effect in time domain, is used for further constrain. Beside the reflections along the *b* axis, the similar constrain for reflections along the *c* axis and the diagonal direction are involved in searching an overall best fit to the intensity changes of all experimentally observed

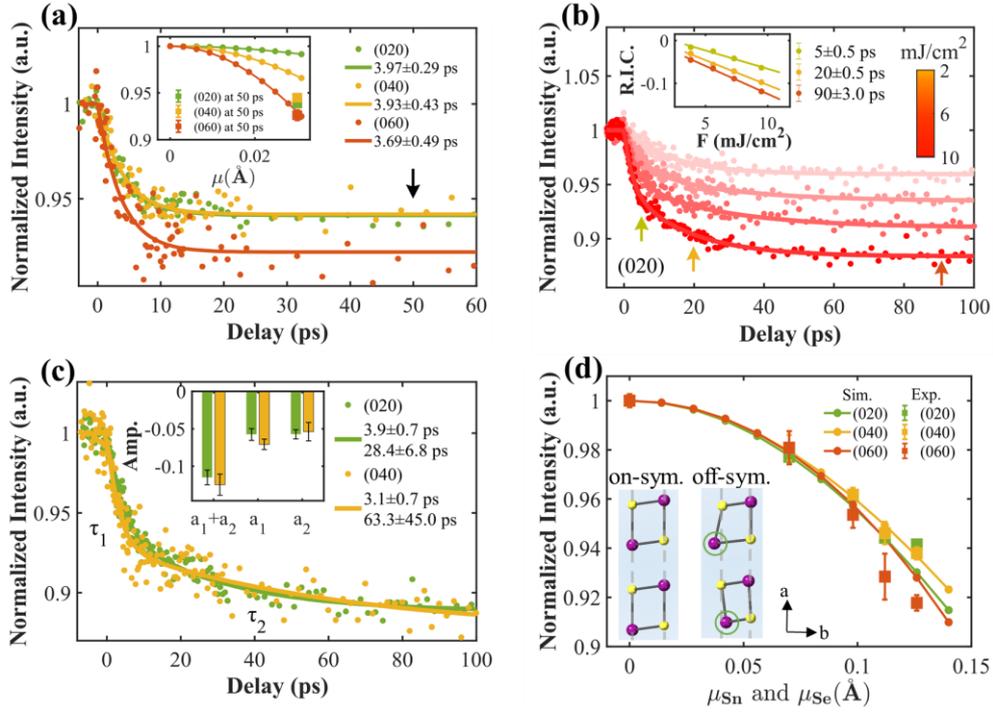

**Figure 2. Identification of the static local disorder projected on the crystallographic *b* axis ((0*k*0), *k* = 2, 4, 6) in the *Pnma* phase of SnSe. (a)** Intensity changes of Bragg reflections (0*k*0) as a function of the time delay with the pump fluence of 7.3 mJ/cm$^2$. The solid lines represent the fit with single exponential function. In the inset, the dotted lines are calculated intensity changes as a function of the thermal vibration amplitude, and the solid squares with error bar are experimental intensity change at the delay time of 50 ps as marked by the black arrow. **(b)** Temporal evolution of the intensity of the (020) reflection at different pump fluences. The inset is, as a function of the pump fluences, the relative intensity change (R.I.C.) at the time delay of 5 ps, 20 ps and 90 ps as marked by arrows in the main panel, and the solid lines represent the linear fit. **(c)** Bi-exponential intensity decay of the (020) and (040) reflection at the pump fluence of 9.5 mJ/cm$^2$, with a bi-exponential fit ($\tau_1$ and $\tau_2$). The amplitudes of the fitting ($a_1$ and $a_2$) are compared in the inset. **(d)** Calculated intensity changes with local disordered structure as a function of the thermal vibration amplitude ($\mu_{Sn}$ and $\mu_{Se}$). The solid squares are experimental intensities at the time delay of -1±1 ps, 2±0.5 ps, 5±0.5 ps, 10±1.0 ps and 30±5.0 ps. The inset displays the structure of the local disorder (off-sym.) used in the calculation, and the rigid crystal structure (on-sym.). The dominant off-symmetry Sn atoms are labeled by green circles. The gray dash lines represent the position of the fraction coordinate 1/4 and 3/4.

Bragg reflections. Finally, 8 local structures with dominant static off-symmetry displacements of Sn and the Debye-Waller effect are ascertained (see Supplementary Table S5.1 and S5.2), each of which may cover one to a few unit cells. Based on the first local structure in Supplementary Table S5.1, Fig. 2d displays the excellent agreement between the experimental and calculated intensity changes over reflections along the *b* axis, as a function of thermal vibration amplitudes. The inset of Fig. 2d is the contrast of the rigid unit cell and the local disordered structure used in the calculation. The main characteristic of this static local disordered structure is the two Sn atoms marked by green circles moving away from the high symmetry position (gray dash lines) along opposite directions. Since the averaged crystalline structure is the *Pnma* phase without any anomalous in the static diffraction pattern, we term the identified multiple local structures with static off-symmetry displacements as static local disorder.

Besides the static local disorder along the *b* axis, the crystal structure projected on the *c* axis also displays local disorders. Fig. 3a displays the time-resolved intensity changes of the (00*l*) (*l* = 2, 4, 6) reflection along the *c* axis. The exponential intensity decay with a time constant of ~5.9 ps indicates a single physical process after photoexcitation. The relative intensity changes over the three reflections at different time delays are summarized in the inset of Fig. 3a. The amplitude of these intensity change increase monotonously as a function of the time delay, suggesting a consistent physical process dominating the intensity changes.

In contrast to the high symmetry crystal structure along the *b* axis, the *c* axis contains a puckered structural distortion in the *Pnma* phase of SnSe, as shown in Fig. 1a. As rising the temperature to above 807 K, the structural transition from the *Pnma* to *Cmcm* phase [22, 24, 26] gives rise to a significant atomic displacements along the *c* axis. Therefore, a possible photoinduced structural transition [55, 56] needs to be evaluated. Fig. 3b displays the calculate the intensity change of the (00*l*) reflections with the structural transition to the *Cmcm* phase and *Immm* phase [55]. Both structural transitions cannot reproduce the experimental intensity changes. Moreover, the calculated temperature rising of 234 K after photoexcitation is far below the temperature required for the transition to the *Cmcm* phase (see Supplementary S2). Therefore, the photoinduced structural transition to the high symmetry phases can be excluded. Similar to the analysis of the *b* axis, we also exclude other possible mechanisms dominating the time-resolved intensity changes along the *c* axis in the rigid *Pnma* phase of SnSe (see Supplementary S5.1 and S5.2). Consequently, static local off-symmetry displacements along the *c* axis need to be introduced into the *Pnma* phase. The consistent physical process, based on the exponential intensity decay in Fig. 3a and the pump fluence dependence in Supplementary S5.3, is thereby a Debye-Waller effect. Then we quantify the static local off-symmetry displacements and the Debye-Waller effect by intensity calculation. The best fit between the time-resolved experimental intensity changes and the calculated intensity changes over all experimentally observed Bragg reflections along the *b* axis, *c* axis and the diagonal direction has ascertained 8 local structures as shown in Table S5.1 and S5.2. With the first local structure in Table S5.2, the calculated intensity changes of the reflections along the *c* axis as a function of thermal vibration amplitudes ($\mu_{Sn}$ and $\mu_{Se}$) display an excellent agreement with the experimental results, as depicted in Fig. 3c. The inset of Fig. 3c is the contrast of the rigid unit cell and the local disordered structure used in the calculation. The main characteristic of this local structure is the two Sn atoms marked by green circles moving away from the high symmetry position (gray dash lines).

Besides that from the reflections along the *b* axis and the *c* axis, the model of the local disorder is further constrained by the reflection (0*ii*) (*i* = 1, 2, 3, 4, 5) along the diagonal direction. In detail, searching the best fit along the *b* axis and the *c* axis respectively provides two libraries containing the local disordered structure and the Debye-Waller effect. The time-resolved intensity change of the diagonal reflections (0*ii*) is determined by both crystallographic axes; therefore, we search the best fit to the time-resolved experimental intensity changes of (0*ii*) with the two identified libraries (see details in Supplementary S6). With the best fit, we finally confirm 8 local structures with static off-symmetry displacements and the Debye-Waller effect, as shown in Supplementary Table S5.1, Table S5.2 and Table S5.3. The experimentally time-resolved intensity changes of the (0*ii*) reflections with a single exponential fit are shown in Fig. 3d. Together with the linear increase of the relative intensity change with rising pump fluence as displayed in the inset of Fig. 3d, a Debye-Waller effect dominated physical process is ascertained, in line with that of the *b* axis and the *c* axis. Fig. 3e displays the good agreement between the calculated and the experimental intensity changes for the (0*ii*) reflections, with the first local structure in Supplementary Table S5.1 and S5.2 and the first thermal vibration amplitude in S5.3. In Fig. 3f, the bar chart shows an excellent agreement between the experimental intensity changes of all acquired Bragg reflections at 60 ps and calculated intensity changes with the local disordered structures and the Debye-Waller effect. At the low temperature of ~90 K, the local disorder and the Debye-Waller effect dominated structural dynamics along the three axes are also confirmed (see Supplementary S7). With the DFT simulation in Supplementary S10, the local disorder by substitution of Sn position with vacancy qualitatively agrees with the experimentally identified local disorder.

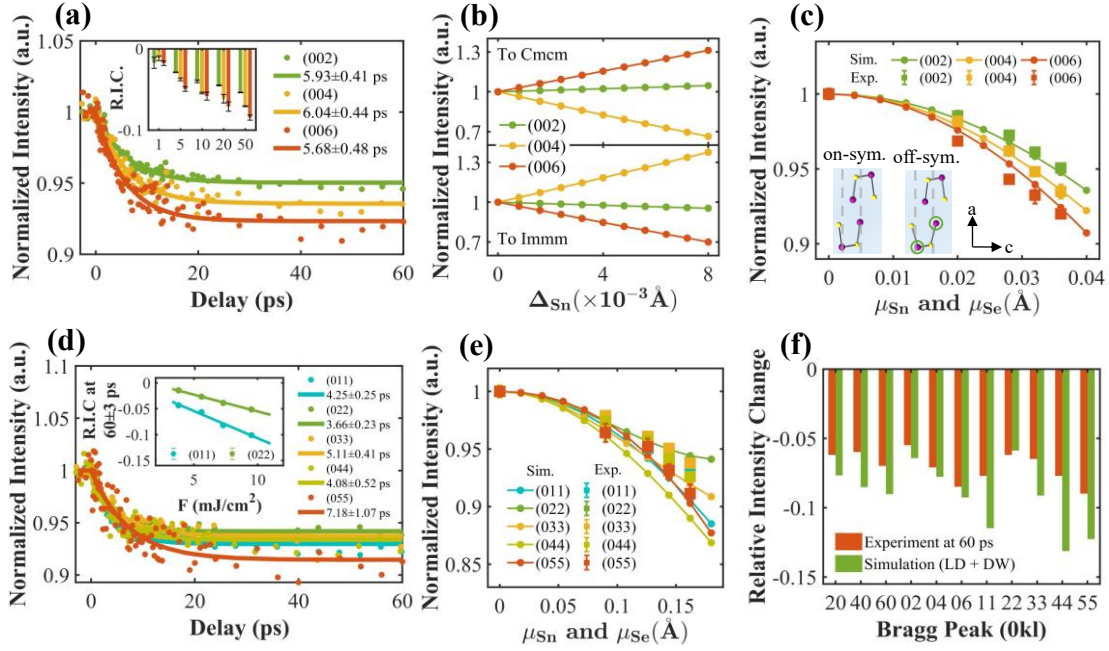

**Figure 3. Identification of the static local disorder projected on the crystallographic c axis ((00$l$), $l$ = 2, 4, 6) and diagonal axis ((0$ii$), $i$ = 1, 2, 3, 4, 5) in the *Pnma* phase of SnSe. (a)** Time resolved intensity changes of (00$l$) reflections along the $c$ axis. The inset is the relative intensity change (R.I.C.) at different time delays (ps). **(b)** Calculated intensity changes with the phase transition to *Cmcm* phase or *Immm* phase. **(c)** Calculated intensity changes with local disordered crystal structure as a function of the thermal vibration amplitude ($\mu_{Sn}$ and $\mu_{Se}$). The solid squares represent the experimental intensity changes at the time delay of -1±1 ps, 2±0.5 ps, 5±0.5 ps, 10±1.0 ps and 30±5.0 ps. The inset displays the structure of the local disorder (off-sym.) used in the calculation, and the rigid crystal structure (on-sym.). The dominant off-symmetry Sn atoms are labeled by green circles. The gray dash lines represent the position of Sn atoms in rigid *Pnma* phase. **(d)** Intensity changes of (0$ii$) reflections along the diagonal axis as a function of the time delay. The inset displays R.I.C. of the (011) and (022) reflection at 60 ps with increasing pump fluence and solid lines are the linear fitting. **(e)** Calculated intensity changes of (0$ii$) reflections with local disordered crystal structure as a function of the thermal vibration amplitude ($\mu_{Sn}$ and $\mu_{Se}$). The solid squares represent the experimental intensity changes at -1±1 ps, 2±0.5 ps , 5±0.5 ps, 10±1.0 ps and 30±5.0 ps. **(f)** Bar chart showing the experimental intensity change at 60 ps and the simulated intensity change with the local disorder (LD) and the Debye-Waller effect (DW).

## 3. Identification of the anharmonicity of the static local disorder

The identified static local disordered structure in the *Pnma* phase of SnSe indicates an intermediate state between the crystalline and the glassy material phase. Therefore, the glass-like structural character, such as localized vibrational modes determined by the local potential energy surfaces [57, 58], is expected to emerge. In Fig. 4a, the intensity of the (015) reflection increases rapidly with a time constant of ~50 fs and an amplitude of ~7% (see details in Supplementary movie), while the intensity of the (013) and (011) reflection remains unchanged in this ultrafast time scale. We exclude some possible mechanisms which may drive these anomalous intensity changes (see details in Supplementary S8). The last possible mechanism is the photoinduced atomic displacement on the local disordered structure. We search the best fit between the calculated intensity change and the experimental intensity change within 0.3 ps over (013), (015) and (00$l$) ($l$ = 2, 4, 6) reflections. For each local structure in Supplementary S5.2, 13 atomic displacements are ascertained, which do not show any relationship to the structural transition to the *Cmcm* or *Immm* phase. Supplementary Table S6 displays the 13 atomic displacements based on the first local disorder in Supplementary S5.2. The photoinduced local displacement suggests the anharmonicity of the local disordered structure; moreover, the ultrafast local displacements further confirm the existence of the local disordered structure. Fig. 4b (left) shows the structure contrast between the rigid crystal structure, the first local disorder in Supplementary S5.2 and the first local displacement from Supplementary Table S6. In the same case, Fig. 4c displays the great agreement between the experimental and the calculated intensity changes of aforementioned reflections with uniformly increasing the amplitude of the local displacement.

To further identify the ultrafast sub-picosecond structural response, we performed the time-dependent density functional theory molecular dynamic (TDDFT-MD) simulations on the rigid *Pnma* phase of SnSe. The dominant motion after ultrafast photoexcitation is the interlayer shear displacement towards the *Immm* phase as shown in Fig. 4d, distinct from the experimentally observed anharmonic displacements. Therefore, the local disorder plays a decisive role in determining the ultrafast structural response. The calculated displacement of ~0.1 Å in Fig. 4d is one order of magnitude larger than the maximum experimental displacement [55]. This remarkable discrepancy may derive from the local disorder induced complex energy relaxation channels in real material system. The local anharmonicity manifests the instability of the local structure with determinate local oscillation modes, analogous to the boson peak with soft local vibrational modes in glasses [57, 58]. Such localized vibrational modes indicates a direct signature of Einstein oscillators, a thermal conduction channel beyond the phonon picture [32], in crystalline SnSe. The ultrafast intensity change within ~0.1 ps in Fig. 4a is generally one quarter of the oscillation period [59]. Accordingly, the average frequency (energy) of the oscillators is 2.5 THz (10.3 meV), which is comparable to that of the Einstein oscillators in other thermoelectric system revealed by inelastic X-ray scattering [17]. These THz Einstein oscillators scatter the heat-carrying lattice phonons and the anharmonicity of the Einstein oscillators enhance the oscillator-phonon scattering as rising the temperature, both of which lower the thermal conductivity in SnSe. Fig. 4b (right) displays the schematic illustration of the local potential energy surfaces of the Einstein oscillators in the ground state, and the flattening of the local potential energy surfaces by photoexcited ultrafast displacements.

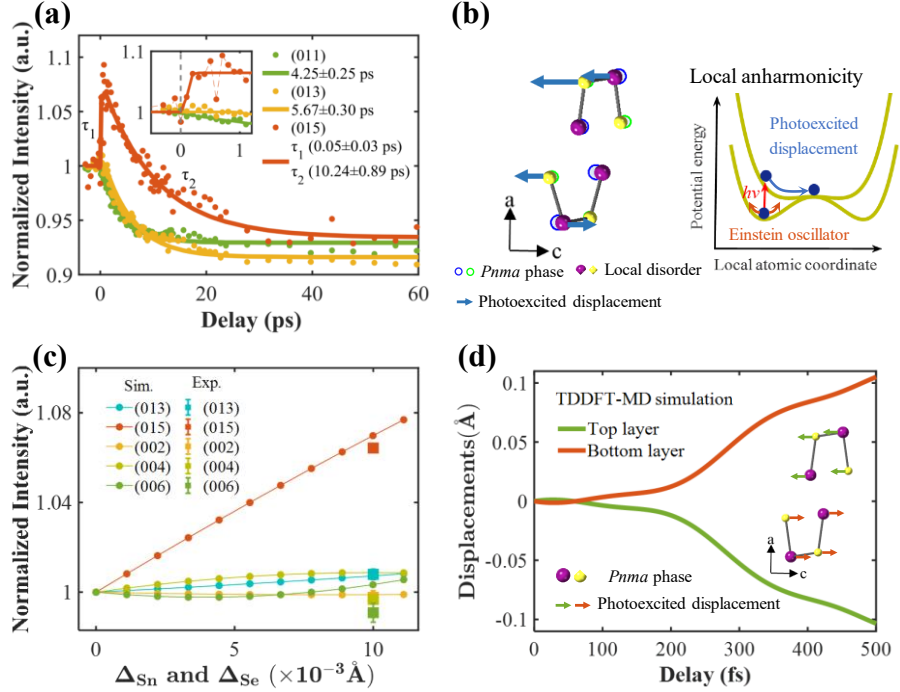

**Figure 4. Identification of the anharmonicity of the local disorder. (a)** Anisotropic intensity changes of Bragg reflections as a function of the time delay. The inset is the temporal evolution within the first 1.2 ps. **(b)** Left: Photoexcited atomic displacements, indicated by arrows, in the local disordered structure. The hollow circles and the solid circles represent the rigid crystal structure and the local disordered structure respectively. Right: Schematic illustration of the local potential energy surfaces of the Einstein oscillators determined by the local disorder, and the photoexcited displacement in the local disordered structure. **(c)** Calculated intensity changes with increasing the displacement of $\Delta_{Sn}$ and $\Delta_{Se}$. The local disordered structure and the directional displacements shown in **(b)** are used in the calculation. The solid squares are the experimental intensity changes at 0.3±0.1 ps. **(d)** Photoexcited interlayer shear displacement in the rigid *Pnma* phase of SnSe by TDDFT-MD simulation. The inset is the schematic illustration of the interlayer shear displacements.

## 4. Discussion and conclusion

With the combination of femtosecond electron diffraction, structure factor calculation and the TDDFT-MD simulation, we reveal the static local disorder and the anharmonicity of the local disorder in thermoelectric SnSe. Determining the local structure is critical since understanding the ultralow thermal conductivity remains phenomenological by fitting simplified scattering models to experimental data [32, 60-62]. The identified local disorder with multiple atomic configurations, which is attributed to the intrinsic vacancy defects in SnSe sample (see details in Supplementary S1 and S10), provides the necessary structural information for quantifying the theoretical thermal transport model [29]. In addition, the 8 degenerate local atomic configurations instead of random configurations evidence local correlated interactions in crystalline disordered materials, such as Pauling's rules [5, 18]. In contrast to the global anharmonicity acrossing the structural transition (*Pnma* to *Cmcm* phase at ~807 K) [24, 26], the anharmonicity of the local disorder, a direct signature of the Einstein oscillators with the energy of ~10.3 meV, is revealed unprecedentedly. The local disorder and local anharmonicity are expected to be the underlying physical mechanism of the mysterious anharmonic behavior measured at near room temperature and below in SnSe [27, 28]. The static local disorder may also play a key role in the unusual in-gap states [63] in SnSe. For crystalline materials with ultralow thermal conductivity, the intermediate regime where both the crystal-like and glass-like thermal transports are associated, is a hot topic in experiment and theory [32, 33] but still in debate [32, 60-62]. A quantitative characterization of the atomic lattice structure and the potential energy landscape is required to resolve the puzzle. In our work, the identified local disorder, the Einstein oscillators and the local anharmonicity directly evidence a glass-like thermal transport channel and update the fundamental insight into the long-debated ultralow thermal conductivity in SnSe. To the best of our knowledge, we first reveal in real space the theoretically predicted THz Einstein oscillators [32]. In addition, the identified anharmonicity of the local disorder is beyond the random hops in conventional glass thermal transport model and we provide the first experimental evidence of such a theoretically predicted thermal transport channel [33, 64]. During preparing our paper, several researches [42, 55] also report the ultrafast structural dynamics of SnSe. Detailed discussion on these researches is shown in Supplementary S11.

In this work, we have identified using femtosecond electron diffraction to reveal local disorder and local anharmonicity in crystalline disordered materials. Our method has several advantages than conventional methods on probing local structure [10, 13-17]. Firstly, the temporal evolution of the local structure is detected and therefore, the static and dynamical local structure can be disentangled, which is hard to achieve previously [13]. Secondly, the anharmonicity of the local structure determined by the local correlated interactions between the phonon, electron and spin subsystem [6-12, 36, 40] can be revealed by ultrafast electronic excitation. Thirdly, the 3D local structure with sub-angstrom displacement can be identified quantitatively by ultrafast structural response, the dimension and the precision of which are much higher than that of the conventional static total scattering and pair distribution function methods [10, 13-17]. Our sophisticated studies in this work call for that, besides the diffuse scattering, the Bragg reflections need to be evaluated in quantifying the local strcutre (see detailes in Supplementary S11). The 2D in plane local disorder in the present work can be simply extended to 3D by additional measuring ultrafast dynamics of the out of plane diffraction spots. The local disorder and local correlated interactions are pervasive and indispensable in a broad of functional materials [5-12, 32, 65]. Therefore, we expect our method will inspire broad interest in material science and ultrafast field.

## Methods

**keV and MeV femtosecond electron diffraction**

The experiments are performed on both the 50 keV and 3 MeV femtosecond electron diffraction system. For the keV femtosecond electron diffraction system, the 800 nm, 35 fs laser and 1 kHz laser pulse (Spectrum Physics Spitfire Ace-35F) is divided into two pulses. One laser pulse used to pump is focused on the sample with a spot size of ~500 μm (FWHM) and an incidence angle around 10° from the sample normal. Another laser pulse is firstly tripled to 266 nm, then focused on the photocathode to generate the probe electron pulse. The spot size of the electron pulse on the sample is ~200 μm (FWHM) focused by a magnetic lens. The diffraction pattern is imaged by a phosphor screen (P43) and recorded by an electron-multiplying charge-coupled device (Andor iXon Ultra 888). The overall temporal resolution of the system is ~500 fs. The base temperature of the sample in experiment is controlled by liquid nitrogen.

For the MeV femtosecond electron diffraction system, the 800 nm, 30 fs and 100 Hz laser pulse (Vitara and Legend Elite Duo HE, Coherent) is split into the pump and the probe pulse. The pump pulse excites the sample and the probe pulse is frequency tripled in nonlinear crystals before illuminating a photocathode for electron pulse generation. After being accelerated by an intense radio-frequency field to relativistic velocity (~0.989c), the electron pulse goes through a double-bend achromatic lens for pulse compression and jitter removal. The spot size of the electron pulse on the sample is ~100 μm FWHM, around five times smaller than the size of the pump laser, ensuring a homogeneous photoexcitation. The diffraction pattern is imaged by a phosphor screen (P43) and recorded by an electron-multiplying charge-coupled device (Andor iXon Ultra 888). The overall temporal resolution of the MeV system is ~50 fs. Further details for the system can be found in ref. [38]. Both the keV and MeV femtosecond electron diffraction experiments give the same time resolved structural dynamics.

**Sample preparation and characterization**

Single crystal SnSe nanofilms are prepared via mechanical exfoliation from the bulk crystal and transferred to TEM grids for femtosecond electron diffraction experiments. The samples with the thickness smaller than 100 nm are characterized by optical microscopy, atomic force microscopy (AFM) and transmission electron microscopy (TEM) (see details in Supplementary S1). We repeat our femtosecond electron diffraction experiments with three different samples and two FED systems (50 keV and 3 MeV FED) to guarantee the reproducible of our experiments and conclusions.

**Structure factor calculation**

The diffraction intensity of the Bragg reflection $I \propto |F|^2$, can be calculated by the structure factor:

$$F(hkl, \Delta t) = \sum_j T_j f_j \cdot exp[-i2\pi(h(x_j + \Delta x_{j,\text{diso}} + \Delta x_{j,\text{disp}}) + k(y_j + \Delta y_{j,\text{diso}} + \Delta y_{j,\text{disp}}) + l(z_j + \Delta z_{j,\text{diso}} + \Delta z_{j,\text{disp}}))]$$

$$T_j = exp(-M), \quad M = 8\pi^2 \langle \mu^2 \rangle (\sin\theta/\lambda)^2$$

See details for the calculation in the main text. As presented in the main text, the local disorder, the Debye-Waller effect and the local anharmonicity are determined by the constrain from the time-resolved intensity changes of the Bragg reflections along the $b$ axis, the $c$ axis and the diagonal direction. The static structure factors without the local disorder calculated by the algebra equation have been compared with that from the SingleCrystal to confirm the reliable structure factor calculation in this work.

**TDDFT-MD and DFT simulation**

The time-dependent density function theory molecular dynamic (TDDFT-MD) calculations are performed in a time-dependent ab initio package (TDAP). Numerical atomic orbitals with double zeta polarization (DZP) are employed as the basis set. The plane-wave energy cutoff is set to 300 Ry. The Brillouin zone is sampled using a 3 × 7 × 7 k-point mesh for the unit cell and the PBE-D3 method of Grimme with Becke-Jonson damping is applied. The Gaussian-type laser pulse utilized in our study is given by $E(t) = E_0 \cos(\omega t) \cdot exp[-(t - t_0)^2 / 2\delta^2]$, where the maximum field intensity $E_0$ is 0.04 V/Å, and $t_0$ = 100 fs is the temporal location of the electric field peak. The laser pulse is linearly polarized along the $y$ axis. The photon energy, and FWHM are set as 1.55 eV ($\lambda$ = 800 nm) and 30 fs, respectively. The laser fluence is 0.17 mJ/cm$^2$. The time-dependent Kohn-Sham wave functions of the system evolve in a micro-canonical ensemble (NVE) ensemble with a time step of 50 attoseconds.

The density function theory (DFT) calculations are performed based on Vienna ab initio simulation package (VASP). The exchange-correlation effect is described within the generalized gradient approximation (GGA) in the Perdew-Burke-Ernzerhof (PBE) functional, together with the projector augmented wave (PAW) potentials. We use a 1 × 2 × 2 supercell with 1 Sn vacancy (Sn$_{15}$Se$_{16}$) in our simulation to verify the disordered structure of SnSe. The Brillouin zone was sampled using a 3 × 5 × 5 k-point mesh for the supercell and the PBE-D3 method of Grimme with Becke-Jonson damping is applied.

## Acknowledgements

We acknowledge Dr. Kunqi Xu and Prof. Zhiwen Shi from School of Physics and Astronomy, Shanghai Jiao Tong University for the characterization of the sample with AFM. **Funding:** This work was supported by National Key R&D Program of China (No.2021YFA1400202), the National Natural Science Foundation of China (11925505), the National Natural Science Foundation of China (11974241), the US-NSF Cooperative agreement (Grant numbers DMR-1644779 and DMR-2128556) and the state of Florida, National Natural Science Foundation of China (No. 12025407, 11934003). **Author Contributions:** Yingpeng Qi devised the project and conceived the presented ideas. Yingpeng Qi and Jingjun Li prepared and characterized the sample. Jingjun Li and Yingpeng Qi performed the keV and MeV femtosecond electron diffraction experiments with help from Luye Yue and Changyuan Yao. Zijing Chen maintained the MeV-UED beamline under the supervision of Dao Xiang. Yingpeng Qi and Jingjun Li analyzed the experimental data and built up the model presented in the paper. Qing Yang performed the TDDFT-MD simulation under the supervision of Sheng Meng. Yingpeng Qi and Jingjun Li wrote the paper with contributions from all other authors. Jianming Cao, Dao Xiang and Sheng Meng supervised the project. **Competing interests:** The authors declare no competing interest. **Data and materials availability:** All data needed to evaluate the conclusions in the paper are present in the paper and/or the Supplementary Materials. Additional data related to this paper may be requested from the corresponding author Yingpeng Qi.

# References

1. M.M. Yang, Z.D. Luo, Z. Mi, J. Zhao, S.P. E, M. Alexe, Piezoelectric and pyroelectric effects induced by interface polar symmetry. *Nature* **584**, 377-381 (2020).
2. J.G. Horstmann, H. Bockmann, B. Wit, F. Kurtz, G. Storeck, C. Ropers, Coherent control of a surface structural phase transition. *Nature* **583**, 232-236 (2020).
3. T. Senthil, Symmetry-protected topological phases of quantum matter. *Annual Review of Condensed Matter Physics* **6**, 299-324 (2015).
4. F. Tang, H.C. Po, A. Vishwanath, X. Wan, Comprehensive search for topological materials using symmetry indicators. *Nature* **566**, 486-489 (2019).
5. A. Simonov, A.L. Goodwin, Designing disorder into crystalline materials. *Nature Reviews Chemistry* **4**, 657-673 (2020).
6. H. Pirie, E. Mascot, C.E. Matt, Y. Liu, P. Chen, M.H. Hamidian, S. Saha, X. Wang, J. Paglione, G. Luke, D. Goldhaber-Gordon, C.F. Hirjibehedin, J.C.S. Davis, D.K. Morr, J.E. Hoffman, Visualizing the atomic-scale origin of metallic behavior in Kondo insulators. *Science* **379**, 1214–1218 (2023).
7. S.H. Skjærvø, Q.N. Meier, M. Feygenson, N.A. Spaldin, S.J.L. Billinge, E.S. Bozin, S.M. Selbach, Unconventional continuous structural disorder at the order-disorder phase transition in the hexagonal manganites. *Physical Review X* **9**, 031001 (2019).
8. E. Dagotto, Complexity in strongly correlated electronic systems. *Science* **309**, 257-262 (2005).
9. E.S. Božin, C.D. Malliakas, P. Souvatzis, T. Proffen, N.A. Spaldin, M.G. Kanatzidis, S.J.L. Billinge, Entropically stabilized local dipole formation in lead chalcogenides. *Science* **330**, 1660-1663 (2010).
10. G.J. Snyder, M. Christensen, E. Nishibori, T. Caillat, B.B. Iversen, Disordered zinc in $Zn_4Sb_3$ with phonon-glass and electron-crystal thermoelectric properties. *Nature Materials* **3**, 458-463 (2004).
11. Y. Qi, N. Chen, T. Vasileiadis, D. Zahn, H. Seiler, X. Li, R. Ernstorfer, Photoinduced ultrafast transition of the local correlated structure in chalcogenide phase-change materials. *Phys Rev Lett* **129**, 135701 (2022).
12. A.V. Kolobov, M. Krbal, P. Fons, J. Tominaga, T. Uruga, Distortion-triggered loss of long-range order in solids with bonding energy hierarchy. *Nat Chem* **3**, 311-316 (2011).
13. S.A.J. Kimber, J. Zhang, C.H. Liang, G.G. Guzman-Verri, P.B. Littlewood, Y. Cheng, D.L. Abernathy, J.M. Hudspeth, Z.Z. Luo, M.G. Kanatzidis, T. Chatterji, A.J. Ramirez-Cuesta, S.J.L. Billinge, Dynamic crystallography reveals spontaneous anisotropy in cubic GeTe. *Nature Materials* **22**, 311-315 (2023).
14. S.J.L. Billinge, I. Levin, The problem with determining atomic structure at the nanoscale. *Science* **316**, 561-565 (2007).
15. J. Shamblin, M. Feygenson, J. Neuefeind, C.L. Tracy, F. Zhang, S. Finkeldei, D. Bosbach, H. Zhou, R.C. Ewing, M. Lang, Probing disorder in isometric pyrochlore and related complex oxides. *Nature Materials* **15**, 507-511 (2016).
16. A. Bencan, E. Oveisi, S. Hashemizadeh, V.K. Veerapandiyan, T. Hoshina, T. Rojac, M. Deluca, G. Drazic, D. Damjanovic, Atomic scale symmetry and polar nanoclusters in the paraelectric phase of ferroelectric materials. *Nat Commun* **12**, 3509 (2021).
17. D.J. Voneshen, K. Refson, E. Borissenko, M. Krisch, A. Bosak, A. Piovano, E. Cemal, M. Enderle, M.J. Gutmann, M. Hoesch, M. Roger, L. Gannon, A.T. Boothroyd, S. Uthayakumar, D.G. Porter, J.P. Goff, Suppression of thermal conductivity by rattling modes in thermoelectric sodium cobaltate. *Nature Materials* **12**, 1028-1032 (2013).
18. E.C. O'Quinn, K.E. Sickafus, R.C. Ewing, G. Baldinozzi, J.C. Neuefeind, M.G. Tucker, A.F. Fuentes, D. Drey, M.K. Lang, Predicting short-range order and correlated phenomena in disordered crystalline materials. *Science Advances* **6**, eabc2758 (2020).
19. Q. Yan, M.G. Kanatzidis, High-performance thermoelectrics and challenges for practical devices. *Nature Materials* **21**, 503-513 (2022).
20. X.L. Shi, J. Zou, Z.G. Chen, Advanced thermoelectric design: From materials and

structures to devices. *Chem Rev* **120**, 7399-7515 (2020).
21. Yu Xiao, L.-D. Zhao, Seeking new, highly effective thermoelectrics. *Science* **367**, 1196-1197 (2020).
22. L.D. Zhao, S.H. Lo, Y. Zhang, H. Sun, G. Tan, C. Uher, C. Wolverton, V.P. Dravid, M.G. Kanatzidis, Ultralow thermal conductivity and high thermoelectric figure of merit in SnSe crystals. *Nature* **508**, 373-377 (2014).
23. L.-D. Zhao, C. Chang, G. Tan, M.G. Kanatzidis, SnSe: A remarkable new thermoelectric material. *Energy & Environmental Science* **9**, 3044-3060 (2016).
24. C.W. Li, J. Hong, A.F. May, D. Bansal, S. Chi, T. Hong, G. Ehlers, O. Delaire, Orbitally driven giant phonon anharmonicity in SnSe. *Nature Physics* **11**, 1063-1069 (2015).
25. Y. Xiao, C. Chang, Y. Pei, D. Wu, K. Peng, X. Zhou, S. Gong, J. He, Y. Zhang, Z. Zeng, L.-D. Zhao, Origin of low thermal conductivity in SnSe. *Physical Review B* **94**, 125203 (2016).
26. T. Lanigan-Atkins, S. Yang, J.L. Niedziela, D. Bansal, A.F. May, A.A. Puretzky, J.Y.Y. Lin, D.M. Pajerowski, T. Hong, S. Chi, G. Ehlers, O. Delaire, Extended anharmonic collapse of phonon dispersions in SnS and SnSe. *Nat Commun* **11**, 4430 (2020).
27. M.Y. Hu, X. Yong, N.J. English, J.S. Tse, Onset of anharmonicity and thermal conductivity in SnSe. *Physical Review B* **104**, 184303 (2021).
28. J.S. Kang, H. Wu, M. Li, Y. Hu, Intrinsic low thermal conductivity and phonon renormalization due to strong anharmonicity of single-crystal tin selenide. *Nano Lett* **19**, 4941-4948 (2019).
29. D. Wu, L. Wu, D. He, L.-D. Zhao, W. Li, M. Wu, M. Jin, J. Xu, J. Jiang, L. Huang, Y. Zhu, M.G. Kanatzidis, J. He, Direct observation of vast off-stoichiometric defects in single crystalline SnSe. *Nano Energy* **35**, 321-330 (2017).
30. L.-D. Zhao, S.-H. Lo, Y. Zhang, H. Sun, G. Tan, C. Uher, C. Wolverton, V.P. Dravid, M.G. Kanatzidis, Zhao et al. Reply. *Nature* **539**, E2-E3 (2016).
31. P.C. Wei, S. Bhattacharya, J. He, S. Neeleshwar, R. Podila, Y.Y. Chen, A.M. Rao, The intrinsic thermal conductivity of SnSe. *Nature* **539**, E1-E2 (2016).
32. S. Mukhopadhyay, D.S. Parker, B.C. Sales, A.A. Puretzky, M.A. McGuire, L. Lindsay, Two-channel model for ultralow thermal conductivity of crystalline $Tl_3VSe_4$. *Science* **360**, 1455–1458 (2018).
33. M. Simoncelli, N. Marzari, F. Mauri, Unified theory of thermal transport in crystals and glasses. *Nature Physics* **15**, 809-813 (2019).
34. Q. Ren, M.K. Gupta, M. Jin, J. Ding, J. Wu, Z. Chen, S. Lin, O. Fabelo, J.A. Rodriguez-Velamazan, M. Kofu, K. Nakajima, M. Wolf, F. Zhu, J. Wang, Z. Cheng, G. Wang, X. Tong, Y. Pei, O. Delaire, J. Ma, Extreme phonon anharmonicity underpins superionic diffusion and ultralow thermal conductivity in argyrodite $Ag_8SnSe_6$. *Nature Materials* (2023).
35. A. Hirata, P. Guan, T. Fujita, Y. Hirotsu, A. Inoue, A.R. Yavari, T. Sakurai, M. Chen, Direct observation of local atomic order in a metallic glass. *Nature Materials* **10**, 28-33 (2011).
36. Y. Qi, M. Guan, D. Zahn, T. Vasileiadis, H. Seiler, Y.W. Windsor, H. Zhao, S. Meng, R. Ernstorfer, Traversing double-well potential energy surfaces: Photoinduced concurrent intralayer and interlayer structural transitions in $XTe_2$ (X = Mo, W). *ACS Nano* **16**, 11124−11135 (2022).
37. Y. Cheng, A. Zong, J. Li, W. Xia, S. Duan, W. Zhao, Y. Li, F. Qi, J. Wu, L. Zhao, P. Zhu, X. Zou, T. Jiang, Y. Guo, L. Yang, D. Qian, W. Zhang, A. Kogar, M.W. Zuerch, D. Xiang, J. Zhang, Light-induced dimension crossover dictated by excitonic correlations. *Nat Commun* **13**, 963 (2022).
38. F. Qi, Z. Ma, L. Zhao, Y. Cheng, W. Jiang, C. Lu, T. Jiang, D. Qian, Z. Wang, W. Zhang, P. Zhu, X. Zou, W. Wan, D. Xiang, J. Zhang, Breaking 50 femtosecond resolution barrier in MeV ultrafast electron diffraction with a double bend achromat compressor. *Phys Rev Lett* **124**, 134803 (2020).
39. M.A. Krivoglaz, X-ray and neutron diffraction in nonideal crystals, 1 ed., Springer Berlin, Heidelberg1996.

40. T. Konstantinova, L. Wu, M. Abeykoon, R.J. Koch, A.F. Wang, R.K. Li, X. Shen, J. Li, J. Tao, I.A. Zaliznyak, C. Petrovic, S.J.L. Billinge, X.J. Wang, E.S. Bozin, Y. Zhu, Photoinduced dynamics of nematic order parameter in FeSe. *Physical Review B* **99**, 180102(R) (2019).
41. B. Guzelturk, T. Winkler, T.W.J. Van de Goor, M.D. Smith, S.A. Bourelle, S. Feldmann, M. Trigo, S.W. Teitelbaum, H.G. Steinruck, G.A. de la Pena, R. Alonso-Mori, D. Zhu, T. Sato, H.I. Karunadasa, M.F. Toney, F. Deschler, A.M. Lindenberg, Visualization of dynamic polaronic strain fields in hybrid lead halide perovskites. *Nature Materials* **20**, 618-623 (2021).
42. L.P. Rene de Cotret, M.R. Otto, J.H. Pohls, Z. Luo, M.G. Kanatzidis, B.J. Siwick, Direct visualization of polaron formation in the thermoelectric SnSe. *Proc Natl Acad Sci U S A* **119**, e2113967119 (2022).
43. L. Waldecker, R. Bertoni, H. Hubener, T. Brumme, T. Vasileiadis, D. Zahn, A. Rubio, R. Ernstorfer, Momentum-resolved view of electron-phonon coupling in multilayer $WSe_2$. *Phys Rev Lett* **119**, 036803 (2017).
44. M. Harb, H. Enquist, A. Jurgilaitis, F.T. Tuyakova, A.N. Obraztsov, J. Larsson, Phonon-phonon interactions in photoexcited graphite studied by ultrafast electron diffraction. *Physical Review B* **93**, 104104 (2016).
45. M. Harb, W. Peng, G. Sciaini, C.T. Hebeisen, R. Ernstorfer, M.A. Eriksson, M.G. Lagally, S.G. Kruglik, R.J.D. Miller, Excitation of longitudinal and transverse coherent acoustic phonons in nanometer free-standing films of (001) Si. *Physical Review B* **79**, 094301 (2009).
46. I. González Vallejo, G. Gallé, B. Arnaud, S.A. Scott, M.G. Lagally, D. Boschetto, P.-E. Coulon, G. Rizza, F. Houdellier, D. Le Bolloc'h, J. Faure, Observation of large multiple scattering effects in ultrafast electron diffraction on monocrystalline silicon. *Physical Review B* **97**, 054302 (2018).
47. H. Zhang, W. Li, J. Essman, C. Quarti, I. Metcalf, W.-Y. Chiang, S. Sidhik, J. Hou, A. Fehr, A. Attar, M.-F. Lin, A. Britz, X. Shen, S. Link, X. Wang, U. Bergmann, M.G. Kanatzidis, C. Katan, J. Even, J.-C. Blancon, A.D. Mohite, Ultrafast relaxation of lattice distortion in two-dimensional perovskites. *Nature Physics* 1-8 (2023).
48. X.-L. Shi, W.-Y. Chen, X. Tao, J. Zou, Z.-G. Chen, Rational structural design and manipulation advance SnSe thermoelectrics. *Materials Horizons* **7**, 3065-3096 (2020).
49. Z. Wang, C. Fan, Z. Shen, C. Hua, Q. Hu, F. Sheng, Y. Lu, H. Fang, Z. Qiu, J. Lu, Z. Liu, W. Liu, Y. Huang, Z.A. Xu, D.W. Shen, Y. Zheng, Defects controlled hole doping and multivalley transport in SnSe single crystals. *Nat Commun* **9**, 47 (2018).
50. J. Li, J. Li, K. Sun, L. Wu, R. Li, J. Yang, X. Shen, X. Wang, H. Luo, R.J. Cava, I.K. Robinson, X. Jin, W. Yin, Y. Zhu, J. Tao, Concurrent probing of electron-lattice dephasing induced by photoexcitation in 1$T$-TaSeTe using ultrafast electron diffraction. *Physical Review B* **101**, 100304(R) (2020).
51. M. Eichberger, H. Schafer, M. Krumova, M. Beyer, J. Demsar, H. Berger, G. Moriena, G. Sciaini, R.J. Miller, Snapshots of cooperative atomic motions in the optical suppression of charge density waves. *Nature* **468**, 799-802 (2010).
52. Vance R. Morrison, Robert. P. Chatelain, Kunal L. Tiwari, Ali Hendaoui, Andrew Bruhács, Mohamed Chaker, B.J. Siwick, A photoinduced metal-like phase of monoclinic $VO_2$ revealed by ultrafast electron diffraction. *Science* **346**, 445-448 (2014).
53. S. Sadasivam, M.K.Y. Chan, P. Darancet, Theory of thermal relaxation of electrons in semiconductors. *Phys Rev Lett* **119**, 136602 (2017).
54. R.P. Chatelain, V.R. Morrison, B.L.M. Klarenaar, B.J. Siwick, Coherent and incoherent electron-phonon coupling in graphite observed with radio-frequency compressed ultrafast electron diffraction. *Physical Review Letters* **113**, 235502 (2014).
55. Y. Huang, S. Yang, S. Teitelbaum, G. De la Peña, T. Sato, M. Chollet, D. Zhu, J.L. Niedziela, D. Bansal, A.F. May, A.M. Lindenberg, O. Delaire, D.A. Reis, M. Trigo, Observation of a novel lattice instability in ultrafast photoexcited SnSe. *Physical Review X* **12**, 011029 (2022).
56. W. Wang, L. Wu, J. Li, N. Aryal, X. Jin, Y. Liu, M. Fedurin, M. Babzien, R. Kupfer, M.

Palmer, C. Petrovic, W. Yin, M.P.M. Dean, I.K. Robinson, J. Tao, Y. Zhu, Photoinduced anisotropic lattice dynamic response and domain formation in thermoelectric SnSe. *npj Quantum Materials* **6**, 97 (2021).
57. T.S. Grigera, V. Martín-Mayor, G. Parisi, P. Verrocchio, Phonon interpretation of the 'boson peak' in supercooled liquids. *Nature* **422**, 289-292 (2003).
58. H. Shintani, H. Tanaka, Universal link between the boson peak and transverse phonons in glass. *Nature Materials* **7**, 870-877 (2008).
59. G. Sciaini, M. Harb, S.G. Kruglik, T. Payer, C.T. Hebeisen, F.J. zu Heringdorf, M. Yamaguchi, M. Horn-von Hoegen, R. Ernstorfer, R.J. Miller, Electronic acceleration of atomic motions and disordering in bismuth. *Nature* **458**, 56-59 (2009).
60. Y. Xia, K. Pal, J. He, V. Ozolins, C. Wolverton, Particlelike phonon propagation dominates ultralow lattice thermal conductivity in crystalline $Tl_3VSe_4$. *Phys Rev Lett* **124**, 065901 (2020).
61. A. Jain, Multichannel thermal transport in crystalline $Tl_3VSe_4$. *Physical Review B* **102**, 201201(R) (2020).
62. Z. Zeng, C. Zhang, Y. Xia, Z. Fan, C. Wolverton, Y. Chen, Nonperturbative phonon scatterings and the two-channel thermal transport in $Tl_3VSe_4$. *Physical Review B* **103**, 224307 (2021).
63. M. Okawa, Y. Akabane, M. Maeda, G. Tan, L.-D. Zhao, M.G. Kanatzidis, T. Suzuki, M. Watanabe, J. Xu, Q. Ren, M. Fujisawa, T. Kanai, J. Itatani, S. Shin, K. Okazaki, N.L. Saini, T. Mizokawa, Direct observation of multiple conduction-band minima in high-performance thermoelectric SnSe. *Scripta Materialia* **223**, 115081 (2023).
64. T. Ghosh, M. Dutta, D. Sarkar, K. Biswas, Insights into low thermal conductivity in inorganic materials for thermoelectrics. *J Am Chem Soc* **144**, 10099-10118 (2022).
65. H. Xie, E.S. Bozin, Z. Li, M. Abeykoon, S. Banerjee, J.P. Male, G.J. Snyder, C. Wolverton, S.J.L. Billinge, M.G. Kanatzidis, Hidden local symmetry breaking in silver diamondoid compounds is root cause of ultralow thermal conductivity. *Adv Mater* **34**, e2202255 (2022).